\documentclass[aps,prb,twocolumn,a4paper,showpacs,eqsecnum,floatfix,fleqn]{revtex4}
\usepackage{amsfonts}
\usepackage{amsmath}
\usepackage{bm}
\usepackage{natbib}
\usepackage{graphicx}
\usepackage{array}
\usepackage{dcolumn}

\begin{document}

\title{A study of the bosonic sector of the two-dimensional \\
Hubbard model within a two-pole approximation}
\author{Adolfo Avella}
\email[E-mail: ]{avella@sa.infn.it}
\author{Ferdinando Mancini}
\email[E-mail: ]{mancini@sa.infn.it} \author{Volodymyr Turkowski}
\altaffiliation[Current address: ]{CFIF, Instituto Superior
Tecnico Av. Rovisco Pais, 1049-001 Lisbon, Portugal}
\affiliation{Dipartimento di Fisica ``E.R. Caianiello'' - Unit\`a INFM di Salerno\\
Universit\`{a} degli Studi di Salerno, I-84081 Baronissi (SA),
Italy}
\date{\today}

\begin{abstract}
The charge and spin dynamics of the two-dimensional Hubbard model
in the paramagnetic phase is first studied by means of the
two-pole approximation within the framework of the Composite
Operator Method. The fully self-consistent scheme requires: no
decoupling, the fulfillment of both Pauli principle and
hydrodynamics constraints, the simultaneous solution of fermionic
and bosonic sectors and a very rich momentum dependence of the
response functions. The temperature and momentum dependencies, as
well as the dependency on the Coulomb repulsion strength and the
filling, of the calculated charge and spin susceptibilities and
correlation functions are in very good agreement with the
numerical calculations present in the literature.
\end{abstract}

 \maketitle

\section{Introduction}

The Hubbard
model\citep{Hubbard:63,Hubbard:64,Hubbard:64a,Hubbard:65} is
capable to describe a rich variety of behaviors including a wide
range of different spin and charge dynamics\citep{Fulde:95}. In
particular, its interactions are thought to be responsible for
strong antiferromagnetic correlations at half-filling and low
temperatures\citep{Dagotto:94}. In the presence of doping, the
antiferromagnetic correlations remain rather strong although the
correlation length can get smaller and smaller on increasing the
doping. The possibility of charge order and phase separation has
also been widely investigated according to the fact that one of
the mostly used derivative of the Hubbard model, the t-J model,
seems to present charge separation for a wide range of external
parameters\citep{Dagotto:94}. However, recent numerical results
seems to indicate that the two models may have different behavior
as far as charge correlations are concerned\citep{Becca:00}.

In this manuscript, we first give a fully self-consistent
treatment of the charge and spin dynamics of the two-dimensional
Hubbard model in the two-pole approximation within the framework
of the Composite Operator Method
(\emph{COM})\citep{Avella:98,Avella:00b}. The fermionic and
bosonic sectors are solved together self-consistently, no
decoupling approximation is used and the explicit momentum
dependence of the spectra involves third nearest-neighbor sites
that forces a rather complex and rich momentum dependence in all
physical properties.

The \emph{COM} rightfully belongs to the family of the projection
methods\citep{Mori:65,Mori:65a,Rowe:68,Roth:69,Nolting:72,
Tserkovnikov:81,Nolting:89,Plakida:89,Fedro:92,Ishihara:94,
Fulde:95,Mancini:95,Mancini:95a,Mancini:95b,Matsumoto:96,Matsumoto:97}
and is based on two main ideas\citep{Mancini:00}: strongly
interacting systems should be described in terms of the
quasi-particles generated by the interactions and the dynamics
should be bounded to the right Hilbert space through the
imposition of constraints coming from the Pauli principle. By
Pauli principle we mean all the relations among operators dictated
by the algebra\citep{Mancini:00}. With respect to other projection
methods the \emph{COM} has some distinguishable peculiarities. In
particular, within the \emph{COM}, there is an absolute freedom to
choose, as asymptotic fields, those that are most suitable with
respect to the properties of the system we wish to describe. This
means that we are not bound to any specific recipe to choose them
and that we can use this freedom to exploit at will the benefits
coming from one choice or another. We can reproduce the results of
the other methods in an unique framework and also go well beyond.
For instance, by choosing suitable asymptotic fields and the
closure of their equations of motion, we were able to describe the
lowest energy scale, which is not algebraic in the model
parameters, of impurity models\citep{Villani:00,Avella:02}. This
result is absolutely precluded to other projection methods that
uniquely focus on the preservation of spectral moments of higher
and higher order. According to this, the method is continuously
developing as we are constantly seeking, one system after the
other, both the most suitable asymptotic fields and the most
effective procedures to determine self-consistently the dynamics.

Once the fermionic propagator is known there are several ways to
compute the response functions (i.e., the retarded propagators of
the two-particle excitations: charge, spin, pair, \ldots). Many
techniques are related to the possible diagrammatic expansions of
the two-particle propagators in terms of the single-particle one
(i.e., the fermionic propagator). However, when operators with
non-canonical commutations are involved the only feasible approach
is based on the one-loop approximation. The complicated algebra of
the composite operators invalidates the Wick theorem and,
consequently, does not allow any simple extension of decoupling
schemes and more involved diagrammatic
approximations\cite{Matsumoto:85,Izyumov:90}. According to this,
we have developed and widely applied a standard procedure to use,
by means of the equations of motion approach, the one-loop
approximation for composite operators\cite{Mancini:95b}.

In this manuscript, we consider another way to tackle the problem:
the two-particle excitations can be considered as a new sector in
the dynamics of the system and we can choose also for them a
suitable asymptotic basis alike it has been done for the fermions.
This gives a new set of equations obeyed by the two-particle
Green's functions and the appearance of zero-frequency constants
and unknown correlators. Also in this case, the enforcement of the
constraints deriving from the Pauli principle allows to compute
all the parameters and to fix the representation of the Hilbert
space\cite{Mancini:00}.

Within the framework of the \emph{COM}, both methods have
advantages and disadvantages. The one-loop approximation becomes
exact in the non-interacting limit, well describes the incoherent
behavior of the two-particle propagators and establishes a tight
connection between the one- and two- particle propagators. These
are really relevant properties once we wish to describe the
bosonic excitations starting from its fundamental constituents:
the electrons. The Fermi surface bending and nesting and the
position of the van Hove singularities strongly influence the
charge and spin response functions. According to this, we managed
to give an explanation for the spin magnetic incommensurability
issue\cite{Mancini:98c} and the overdoped transition of the
cuprate superconductors\cite{Avella:98c}. On the other hand, the
one-loop approximation is not adequate to describe the system in
the proximity of ordered phases as the dynamics of the bosonic
excitations is mainly described in terms of \emph{scattering} of
elementary electronic excitations. This practically induces so
strong finite life-time effects to prevent any possible softening
of the bosonic modes. As discussed in the above, any possible
extension involves so complicated diagrammatic expansions to be
practically unfeasible.

As regards the pole approximation for the two-particle propagators
we have obvious advantages like: the possibility to easily get the
spectra and the analytical expressions of correlation functions
and susceptibilities; the capability to study instabilities (i.e.,
the softening of the modes) in the whole range of model and
external parameters; the possibility to consider the bosonic
excitations as the media of new interactions among the electrons.
In this paper, we show that it is possible to get spin
antiferromagnetic correlations and weak charge ordering tendency
at commensurate filling in exceptionally good agreement with the
numerical results present in the literature. On the other hand,
the pole approximation is based on a description of the bosonic
excitations as true quasi-particles: the two-particle properties
are entirely controlled by the dynamics, which is only weakly
\emph{renormalized} by the fermions; the single-particle
properties (e.g., Fermi surface shape, position of the van Hove
singularity, \ldots) do not influence significatively the response
function behaviors; the finite life-time effects are completely
neglected and the tendency towards ordering (i.e., softening) is
sometime exaggerated. Anyway, the use of the Green's function
formalism for the bosonic sector opens the possibility to explore
another really relevant issue: the ergodicity of the bosonic
dynamics and the presence of zero-frequency constants in the
expression of the casual Green's function and of the correlation
functions\cite{Kubo:57,Zubarev:60,Mancini:00}. In this manuscript,
we decided not to pursue this analysis and to fix the
zero-frequency constant values by means of ergodicity requirements
in accordance with the general understanding of bulk systems.

As we can see, the two methods are effectively complementary and
can be used to describe the spin and charge dynamics of the system
in different region of the parameter space according to the
relevance and prevalence of localization and ordering (two-pole)
with respect to delocalization and symmetry (one-loop).

It is also worth noting that the pole approximation allows, at
least in principle, to get a completely self-consistent solution
putting together fermionic, spin, charge and pair
propagators\citep{Mancini:94a}. The Pauli principle could be then
used to get also the zero-frequency constants in self-consistency
and definitely fix the Hilbert space, as described in
Ref.~\onlinecite{Mancini:00}.

\section{The Hubbard model and the fermionic sector}

The Hubbard model is described by the following Hamiltonian
\begin{equation}
H=\sum_{\mathbf{ij}}\left( t_{\mathbf{ij}}-\mu \,\delta
_{\mathbf{ij}}\right) c^{\dagger }\left( i\right) \,c\left(
j\right) +U\sum_{\mathbf{i}}n_{\uparrow }\left( i\right)
\,n_{\downarrow }\left( i\right)
\end{equation}
where $c^{\dagger }\left( i\right) =\left( c_{\uparrow }^{\dagger
}\left( i\right) \, c_{\downarrow }^{\dagger }\left( i\right)
\right) $ is the electronic creation operator in spinorial
notation at the site $\mathbf{i}$ [$i=\left( \mathbf{i},t\right)
$] and $n_{\sigma }\left( i\right) =c_{\sigma }^{\dagger }\left(
i\right) \,c_{\sigma }\left( i\right) $ is the number operator for
spin $\sigma $ at the site $\mathbf{i}$; $\mu $ is the chemical
potential and $U$ is the on-site Coulomb repulsion.

The matrix $t_{\mathbf{ij}}$ describes the nearest-neighbor
hopping; in the 2D case we have $t_{\mathbf{ij}}=-4t\,\alpha
_{\mathbf{ij}}$, where
\begin{equation}
\alpha_{\mathbf{ij}}=\frac{1}{N}\sum_{\mathbf{k}}e^{\mathrm{i}\,\mathbf{k}(\mathbf{i}
-\mathbf{j})}\alpha \left( \mathbf{k}\right)
\end{equation}
is the projector on the nearest-neighbor sites and $\alpha \left(
\mathbf{k} \right) =\frac{1}{2}\left[ \cos \left( k_{x}\,a\right)
+\cos \left( k_{y}\,a\right) \right] $ and $a$ is the lattice
parameter.

We choose the following fermionic
basis\cite{Mancini:95,Mancini:95a,Mancini:95b}
\begin{equation}
\Psi \left( i\right) =\left(
\begin{array}{c}
\xi \left( i\right) \\
\eta \left( i\right)
\end{array}
\right)
\end{equation}
where $\xi \left( i\right) =\left[ 1-n\left( i\right) \right]
c\left( i\right) $ and $\eta \left( i\right) =n\left( i\right)
\,c\left( i\right) $ are the Hubbard operators. $\Psi \left(
i\right) $ satisfies the following equation of motion
\begin{equation}\label{current}
J\left( i\right) =\mathrm{i}\frac{\partial }{\partial t}\Psi
\left( i\right) =\left(
\begin{array}{c}
-\mu \,\xi \left( i\right) -4t\,c^{\alpha }\left( i\right)
-4t\,\pi \left(
i\right) \\
-(\mu -U)\eta \left( i\right) +4t\,\pi \left( i\right)
\end{array}
\right)
\end{equation}
where $c^{\gamma}\left( \mathbf{i},t\right)
=\sum_{\mathbf{j}}\gamma_{\mathbf{ij}}\,c\left(
\mathbf{j},t\right) $ [$\gamma_{\mathbf{ij}}$ stands for any
projector on the $\mathbf{j}$ neighbor sites of $\mathbf{i}$; see
Appendix] and $\pi \left( i\right) =\frac{1}{2}\sigma ^{\mu
}\,n_{\mu }\left( i\right) \,c^{\alpha }\left( i\right) +c\left(
i\right) \left[ c^{\dagger \alpha }\left( i\right) \,c\left(
i\right) \right] $. $n_{\mu }(i)=c^{\dagger }(i)\,\sigma _{\mu
}\,c(i)$ are the charge ($\mu =0$) and spin ($\mu =1,2,3$ )
density operators, with $\sigma _{\mu }=\left(
1,\vec{\sigma}\right) $, $ \sigma ^{\mu }=\left(
-1,\vec{\sigma}\right) $ and $\vec{\sigma}$ are the Pauli
matrices.

Let us project the source $J\left( i\right)$ on the chosen basis
\begin{equation}
J\left( \mathbf{i},t\right) \cong \sum_{\mathbf{j}}\varepsilon
\left( \mathbf{i},\mathbf{j}\right) \,\Psi \left(
\mathbf{j},t\right) \label{psipem}
\end{equation}
Accordingly, the energy matrix $\varepsilon \left(
\mathbf{i},\mathbf{j}\right) $ is defined through the equation
\begin{equation}
m\left( \mathbf{i},\mathbf{j}\right) =\sum_{\mathbf{l}}\varepsilon
\left( \mathbf{i},\mathbf{l}\right) \,I\left(
\mathbf{l},\mathbf{j}\right)
\end{equation}
where the $m$-matrix and the normalization matrix $I$ have the
following definitions
\begin{align}
&m\left( \mathbf{i},\mathbf{j}\right) =\left\langle \left\{
J\left( \mathbf{i},t\right) ,\Psi
^{\dagger }\left( \mathbf{j},t\right) \right\} \right\rangle \label{curr1}\\
&I\left( \mathbf{i},\mathbf{j}\right) =\left\langle \left\{ \Psi
\left( \mathbf{i},t\right) ,\Psi ^{\dagger }\left(
\mathbf{j},t\right) \right\} \right\rangle
\end{align}
It is worth pointing out that in Eq.~(\ref{curr1}) $J(i)$ is the
total current given in Eq.~(\ref{current}) and not the
approximated one. After Eq.~(\ref{psipem}), the Fourier transform
of the thermal single-particle retarded Green's function $G\left(
i,j\right) =\langle R\left[ \Psi \left( i\right) \,\Psi ^{\dagger
}\left( j\right) \right] \rangle $ satisfies the following
equation
\begin{equation}
\left[ \omega -\varepsilon \left( \mathbf{k}\right) \right]
G\left( \mathbf{k },\omega\right) =I\left( \mathbf{k}\right)
\end{equation}
The straightforward application of this
scheme\citep{Mancini:95,Mancini:95a,Mancini:95b,Mancini:96} gives
that, in the paramagnetic phase, $I\left( \mathbf{k}\right) $ has
diagonal form with $I_{11}=1-n/2$ and $I_{22}=n/2$ ($ \langle
n_{\sigma }\left( i\right) \rangle =\frac{n}{2}$) and that the
$m$-matrix depends on three parameters: the chemical potential
$\mu $ and two correlators
\begin{align}
&\Delta = \langle \xi ^{\alpha }\left( i\right) \,\xi ^{\dagger
}\left( i\right) \rangle -\langle \eta ^{\alpha }\left( i\right)
\,\eta ^{\dagger
}\left( i\right) \rangle \\
&p = \frac14\langle n_{\mu }^{\alpha }\left( i\right) \,n_{\mu
}\left( i\right) \rangle -\langle \lbrack c_{\uparrow }\left(
i\right) \,c_{\downarrow }\left( i\right) ]^{\alpha }c_{\downarrow
}^{\dagger }\left( i\right) \,c_{\uparrow }^{\dagger }\left(
i\right) \rangle
\end{align}
The three parameters $\mu $, $\Delta $ and $p$ are functions of
the external parameters $n$, $T$ and $U$ and can be determined
self-consistently through a set of three coupled equations
\begin{equation}
\left\{
\begin{array}{l}
n=2\left[ 1-\left\langle c\left( i\right) \,c^{\dagger }\left(
i\right)
\right\rangle \right] \\
\Delta =\langle \xi ^{\alpha }\left( i\right) \,\xi ^{\dagger
}\left( i\right) \rangle -\langle \eta ^{\alpha }\left( i\right)
\,\eta ^{\dagger
}\left( i\right) \rangle \\
\langle \xi \left( i\right) \,\eta ^{\dagger }\left( i\right)
\rangle =0
\end{array}
\right.
\end{equation}
The first equation fixes the particle number, the second one comes
from the definition of $\Delta $ and the third one assures that
the solution respects the Pauli principle (i.e., the local
algebra)\citep{Mancini:00}. In this latter equation resides the
main difference with all the other two-pole approximations. This
equation: allows to fix the representation\cite{Mancini:00};
implements the particle-hole symmetry in the
solution\cite{Avella:98}; avoids uncontrolled decoupling or
further approximations on higher order correlators. Using this
procedure is possible to find two solutions: one with a $p$
positive and of order of the filling $n$ and another with $p$
manly negative and rather small. The main difference between these
two solutions resides in the strength of the antiferromagnetic
correlations\citep{Avella:98e,Avella:00}.

It is worth noting that this set of coupled self-consistent
equations gives the exact solution in the atomic and in the
non-interacting cases as well as for the interacting two-site
system\cite{Avella:01}. According to this, we are confident to
reproduce at least the two most relevant scale of energies in the
system: the Coulomb repulsion $U$ and the exchange energy $J$. The
latter is already well defined on the two-site system that is the
minimal cluster where $J$ becomes effective.

Within this calculation scheme, the fermionic solution is known in
a fully self-consistent manner and without opening the bosonic
sector. Once we have the electronic Green's function we can get
all single-particle, local and thermodynamic properties
straightforwardly. In the last years, by means of the \emph{COM}
in the above described approximation, we got results in excellent
agreement with numerical and exact solutions as regards many
lattice and impurity
systems\cite{Mancini:95,Mancini:96d,Matsumoto:96,Matsumoto:97,Avella:98,Avella:98c,Avella:98d,Avella:98e,
Mancini:98c,Mancini:99c,Mancini:00b,Avella:00b,Villani:00,Fiorentino:01}.

\section{Charge and spin response properties}

As stated in the introduction we choose to compute the charge and
spin response functions by studying the causal Green's
function\citep{Mancini:94a} $G^{(\mu )}\left(
i,j\right)=\left\langle T\left[ n_{\mu }\left( i\right) \,n_{\mu
}\left( j\right) \right] \right\rangle $. As we widely discussed
in Ref.~\onlinecite{Mancini:00}, to obtain a correct description
of the bosonic properties is necessary to compute first the causal
Green's function and then derive from this latter all other
propagators and correlators. The reason of this lies in the fact
that the zero-frequency constants do not explicitly contribute to
the retarded functions, although there is an implicit dependence
through the self-consistent determination of the internal
parameters. Starting from the retarded function would lead to
ignore the zero-frequency constants and will give \textbf{wrong
results}. Once we know the Fourier transform of $G^{(\mu )}\left(
i,j\right)$, that is $G^{(\mu )}\left( \mathbf{k},\omega \right)
$, we can find spin and charge susceptibilities $\chi ^{(\mu
)}\left( \mathbf{k},\omega \right)= -F\left\langle R\left[ n_{\mu
}\left( i\right) \,n_{\mu }\left( j\right) \right] \right\rangle$
and correlation functions $C^{(\mu )}\left( \mathbf{k},\omega
\right)=F\left\langle n_{\mu }\left( i\right) \,n_{\mu }\left(
j\right)\right\rangle $ by means of the well-known formulas
\begin{align}
&\Re \left[ \chi ^{(\mu )}\left( \mathbf{k},\omega \right) \right]
=-\Re \left[ G^{(\mu )}\left( \mathbf{k},\omega \right) \right] \label{Rechi}\\
&\Im \left[ \chi ^{(\mu )}\left( \mathbf{k},\omega \right) \right]
=-\tanh \frac{\omega }{2T}\Im \left[ G^{(\mu )}\left(
\mathbf{k},\omega \right) \right] \label{Imchi}\\
&C^{(\mu )}\left( \mathbf{k},\omega \right) =-\left( 1+\tanh
\frac{\omega }{ 2T}\right) \Im \left[ G^{(\mu )}\left(
\mathbf{k},\omega \right) \right] \label{Corr}
\end{align}
where $T$, $R$ and $F$ are the causal and retarded operators and
the Fourier transform, respectively.

As widely discussed in the introduction, in this manuscript we
will study the spin and charge channels of the bosonic sector by
using a pole approximation. Let us write the equation of motion
for the operator $n_{\mu }\left( i\right) $
\begin{equation}
\mathrm{i}\frac{\partial }{\partial t}n_{\mu }\left( i\right)
=-4t\,\rho _{\mu }\left( i\right)
\end{equation}
where
\begin{equation}
\rho _{\mu }\left( i\right) =c^{\dagger }\left( i\right) \,\sigma
_{\mu }\,c^{\alpha }\left( i\right) -c^{\dagger \alpha }\left(
i\right) \,\sigma _{\mu }\,c\left( i\right)
\end{equation}
The bosonic basis has to be chosen in order to be compatible with
the fermionic one and with a non-local component as we wish to
take into account, at least partially, the delocalization driven
by the kinetic term of the Hamiltonian. According to this, the
most natural choice is a two-component basis and, in particular,
that directly generated by the hierarchy of the equations of
motion. This will assure that the first four bosonic spectral
moments have the correct functional form\citep{Mancini:98b}.
Therefore, we take as bosonic basis the following one
\begin{equation}
N_{\mu }\left( i\right) =\left(
\begin{array}{c}
n_{\mu }\left( i\right) \\
\rho _{\mu }\left( i\right)
\end{array}
\right)
\end{equation}
The equation of motion of $\rho _{\mu }\left( i\right) $ is the
following one
\begin{equation}
\mathrm{i}\frac{\partial }{\partial t}\rho _{\mu }\left( i\right)
=-4t\,l_{\mu }\left( i\right) +U\,\kappa_{\mu }\left( i\right)
\end{equation}
where the higher-order bosonic operators are defined by
\begin{eqnarray}
\kappa_{\mu }\left( i\right) &=&c^{\dagger }\left( i\right)
\,\sigma _{\mu }\,\eta ^{\alpha }\left( i\right) -\eta ^{\dagger
}\left( i\right) \,\sigma
_{\mu }\,c^{\alpha }\left( i\right) \nonumber \\
&&+\eta ^{\dagger \alpha }\left( i\right) \,\sigma _{\mu
}\,c\left( i\right)
-c^{\dagger \alpha }\left( i\right) \,\sigma _{\mu }\,\eta \left( i\right) \\
l_{\mu }\left( i\right) &=&c^{\dagger }\left( i\right) \,\sigma
_{\mu }\,c^{\alpha ^{2}}\left( i\right) +c^{\dagger \alpha
^{2}}\left( i\right)
\,\sigma _{\mu }\,c\left( i\right) \nonumber \\
&&-2c^{\dagger \alpha }\left( i\right) \,\sigma _{\mu }\,c^{\alpha
}\left( i\right)
\end{eqnarray}
and the definition of $c^{\alpha ^{2}}\left( i\right)$  can be
found in Appendix.

Using the same procedure used for the fermions, we have
\begin{equation}
\mathrm{i}\frac{\partial }{\partial t}N_{\mu }\left(
\mathbf{i},t\right) \cong \sum_{\mathbf{j}}\varepsilon ^{(\mu
)}\left( \mathbf{i},\mathbf{j}\right) \,N_{\mu }\left(
\mathbf{j},t\right)
\end{equation}
where $\varepsilon ^{(\mu )}\left( \mathbf{i},\mathbf{j}\right) $
is given by
\begin{equation}
m^{(\mu )}\left( \mathbf{i},\mathbf{j}\right)
=\sum_{\mathbf{l}}\varepsilon ^{(\mu )}\left(
\mathbf{i},\mathbf{l}\right) \,I^{(\mu)}\left(
\mathbf{l},\mathbf{j}\right)
\end{equation}
and the normalization matrix $I^{(\mu)}$ and the
$m^{(\mu)}$-matrix have the following definitions
\begin{eqnarray}
I^{(\mu)}\left( \mathbf{i},\mathbf{j}\right) &=&\left\langle
\left[ N_{\mu }\left( \mathbf{i}
,t\right) ,N_{\mu }^\dagger\left( \mathbf{j},t\right) \right] \right\rangle \\
m^{(\mu )}\left( \mathbf{i},\mathbf{j}\right) &=&\left\langle
\left[ \mathrm{i}\frac{\partial }{\partial t}N_{\mu }\left(
\mathbf{i},t\right) ,N_{\mu }^\dagger\left( \mathbf{j} ,t\right)
\right] \right\rangle
\end{eqnarray}
As it can be easily verified, in the paramagnetic phase the
normalization matrix $I^{(\mu)}$ does not depend on the index
$\mu$; charge and spin operators have the same weight. The two
matrices $I^{(\mu)}$ and $m^{(\mu)}$ have the following form in
momentum space\citep{Mancini:94a}
\begin{equation}
I^{(\mu)}\left( \mathbf{k}\right) =\left(
\begin{array}{cc}
0 & I^{(\mu)}_{12}\left( \mathbf{k}\right) \\
I^{(\mu)}_{12}\left( \mathbf{k}\right) & 0
\end{array}
\right)
\end{equation}
\begin{equation}
m^{(\mu )}\left( \mathbf{k}\right) =\left(
\begin{array}{cc}
m_{11}^{(\mu )}\left( \mathbf{k}\right) & 0 \\
0 & m_{22}^{(\mu )}\left( \mathbf{k}\right)
\end{array}
\right)
\end{equation}
where
\begin{eqnarray}
I^{(\mu)}_{12}\left( \mathbf{k}\right) &=&4\left[ 1-\alpha \left(
\mathbf{k}
\right) \right] C^{\alpha } \\
m_{11}^{(\mu )}\left( \mathbf{k}\right)
&=&-4t\,I^{(\mu)}_{12}\left(
\mathbf{k}\right) \\
m_{22}^{(\mu )}\left( \mathbf{k}\right) &=&-4t\,I_{l_{\mu }\rho
_{\mu }}\left( \mathbf{k}\right) +U\,I_{\kappa_{\mu }\rho _{\mu
}}\left( \mathbf{k} \right)
\end{eqnarray}
The exact expressions of the normalization matrix entries and the
definition of the parameters they depend on can be found in the
Appendix. The energy matrix $\varepsilon^{(\mu )}\left(
\mathbf{k}\right)$ has off-diagonal form with non-zero elements
\begin{eqnarray}
\varepsilon _{12}^{(\mu )}\left( \mathbf{k}\right) &=&-4t \\
\varepsilon _{21}^{(\mu )}\left( \mathbf{k}\right)
&=&\frac{m_{22}^{(\mu )}\left( \mathbf{k}\right)
}{I^{(\mu)}_{12}\left( \mathbf{k}\right) }
\end{eqnarray}

For the sake of simplicity, we will now extend the previous used
notation for the bosonic causal Green's function $G^{(\mu )}\left(
i,j\right)=\left\langle T\left[ n_{\mu }\left( i\right) \,n_{\mu
}\left( j\right) \right] \right\rangle$ to the complete $2\times2$
matricial one, that is, $G^{(\mu )}\left( i,j\right)$ is hereafter
defined as $\left\langle T\left[ N_{\mu }\left( i\right) \,N_{\mu
}^{\dagger}\left( j\right) \right] \right\rangle $. We will also
use the accordingly extended notation for the correlation function
$C^{(\mu )}\left( i,j\right)$. They have then the following
expressions
\begin{multline}
G^{(\mu )}\left( \mathbf{k},\omega \right)=-\mathrm{i}\frac{(2\pi
)^{3}}{a^2}\delta ^{(2)}\left(
\mathbf{k}\right) \,\delta (\omega )\,\Gamma_{\mu} \\
+\sum_{n=1}^{2}\sigma ^{(n,\mu )}\left(
\mathbf{k}\right)\frac{1+f_\textrm{B}(\omega )}{\omega
-\omega _{n}^{(\mu )}\left( \mathbf{k}\right) +\mathrm{i}\,\delta } \\
-\sum_{n=1}^{2}\sigma ^{(n,\mu )}\left( \mathbf{k}\right)\frac{
f_\textrm{B}(\omega )}{\omega -\omega _{n}^{(\mu )}\left(
\mathbf{k}\right) -\mathrm{i}\,\delta }
\end{multline}
\begin{multline}
C^{(\mu )}\left( \mathbf{k},\omega \right) =\frac{(2\pi
)^{3}}{a^2}\delta ^{(2)}\left(
\mathbf{k}\right) \,\delta (\omega )\,\Gamma_{\mu} \\
+2\pi\sum_{n=1}^{2}\delta\left[\omega-\omega _{n}^{(\mu )}\left(
\mathbf{k}\right)\right]\left[1+f_\textrm{B}(\omega )\right]\sigma
^{(n,\mu )}\left( \mathbf{k}\right)
\end{multline}
where $\Gamma_{\mu}$ is the zero-frequency
constant\citep{Mancini:00} and $f_\textrm{B}(\omega
)=\frac{1}{e^{\beta\,\omega}-1}$ is the Bose-Einstein distribution
function. In this manuscript, we will use the ergodic value (i.e.,
$\Gamma_{11\mu}=\delta _{\mu 0}\,n^{2}$) for the zero-frequency
constant as explained in the introduction. The energy spectra are
given by
\begin{eqnarray}
&&\omega _{n}^{(\mu )}\left( \mathbf{k}\right) =\left( -\right)
^{n}\omega^{(\mu )}
\left( \mathbf{k}\right) \\
&& \omega^{(\mu )} \left( \mathbf{k}\right) = \sqrt{ \varepsilon
_{12}^{(\mu )}\left( \mathbf{k}\right) \,\varepsilon _{21}^{(\mu
)}\left( \mathbf{k}\right) }
\end{eqnarray}
and the spectral functions have the following expression
\begin{equation}
\sigma ^{(n,\mu )}\left( \mathbf{k}\right)
=\frac{1}{2}I^{(\mu)}_{12}\left( \mathbf{k}\right) \left(
\begin{array}{cc}
\frac{\varepsilon _{12}^{(\mu )}\left( \mathbf{k}\right) }{\omega
_{n}^{(\mu )}\left( \mathbf{k}\right) } & 1 \\
1 & \frac{\varepsilon _{21}^{(\mu )}\left( \mathbf{k}\right)
}{\omega _{n}^{(\mu )}\left( \mathbf{k}\right) }
\end{array}
\right)
\end{equation}
As it can be seen from the expressions given in Appendix, the
Green's function and the correlation function depend on various
parameters, static correlation functions, that must be
self-consistently calculated. A subset of parameters, $C^{\alpha
}$, $C^{\lambda }$, $ C^{\mu }$, $E^{\beta }$ and $E^{\eta }$, are
of fermionic nature and can be computed through the fermionic
Green's function. The \emph{negative} $p$ solution will be used in
order to get enhanced antiferromagnetic correlations. The
remaining parameters, $ a_{\mu }$, $b_{\mu }$, $c_{\mu }$ and
$d_{\mu }$, are of bosonic nature, but they cannot be expressed in
terms of the bosonic Green's function under study as they belong
to higher order propagators. As in the fermionic sector, we can
avoid studying complicated higher order propagators requiring the
fulfillment of the Pauli principle and of other symmetry
requirements. Four equations will be used to fix these parameters:
one equation comes from the Pauli principle and other three from
the general properties of the bosonic spectra $\omega _{n}^{(\mu
)}\left( \mathbf{k}\right) $ for small momenta (i.e., for
$k\rightarrow 0$ where $k=\sqrt{k_{x}^{2}+k_{y}^{2}}$). The Pauli
principle\citep{Mancini:00} gives
\begin{subequations}
\label{Pauli}
\begin{eqnarray}
\left\langle n\left( i\right) \,n\left( i\right) \right\rangle
&=&n+2D \\
\left\langle n_{k}\left( i\right) \,n_{k}\left( i\right)
\right\rangle &=&n-2D\qquad k=1,\text{ }2,\text{ }3
\end{eqnarray}
where $D=\left\langle n_{\uparrow }\left( i\right) n_{\downarrow
}\left( i\right) \right\rangle =\frac{n}{2}-\left\langle
\eta\left( i\right) \,\eta ^{\dagger }\left( i\right)
\right\rangle$ is the double occupancy. From the continuity
equation\citep{Abrikosov:63} it follows that
\end{subequations}
\begin{equation}
\lim_{k\rightarrow 0}\omega _{n}^{(\mu )}\left( \mathbf{k}\right)
\cong c_{n}^{(\mu )}\,k^{s}
\end{equation}
where $s\geq 1$ and $c_{n}^{(\mu )}$ is the \emph{velocity}. Let
us analyze the expression for $\omega _{n}^{(\mu )}\left(
\mathbf{k}\right) $. The function $m_{22}^{(\mu )}\left(
\mathbf{k} \right) $ at small $k$ can be cast in the following
form
\begin{eqnarray}
m_{22}^{(\mu )}\left( \mathbf{k}\right) &=&m_{0}^{(\mu
)}+m_{1}^{(\mu )}(k\,a)^{2}+m_{2}^{(\mu )}(k\,a)^{4} \nonumber \\
&&+m_{3}^{(\mu )}(k\,a)^{4}\sin ^{2}(2\phi _{k})+O((k\,a)^{6})
\end{eqnarray}
where $\phi _{k}=\arctan \frac{k_{y}}{k_{x}}$. The function
$I^{(\mu)}_{12}\left( \mathbf{k}\right) \ $behaves as
$(k\,a)^{2}C^{\alpha }$ at small $k$. Therefore, to satisfy the
continuity equation we must put
\begin{equation}
m_{0}^{(\mu )}=m_{1}^{(\mu )}=0 \label{hydro}
\end{equation}

\begin{figure}[tbp]
\begin{center}
\includegraphics[width=8cm,keepaspectratio=true]{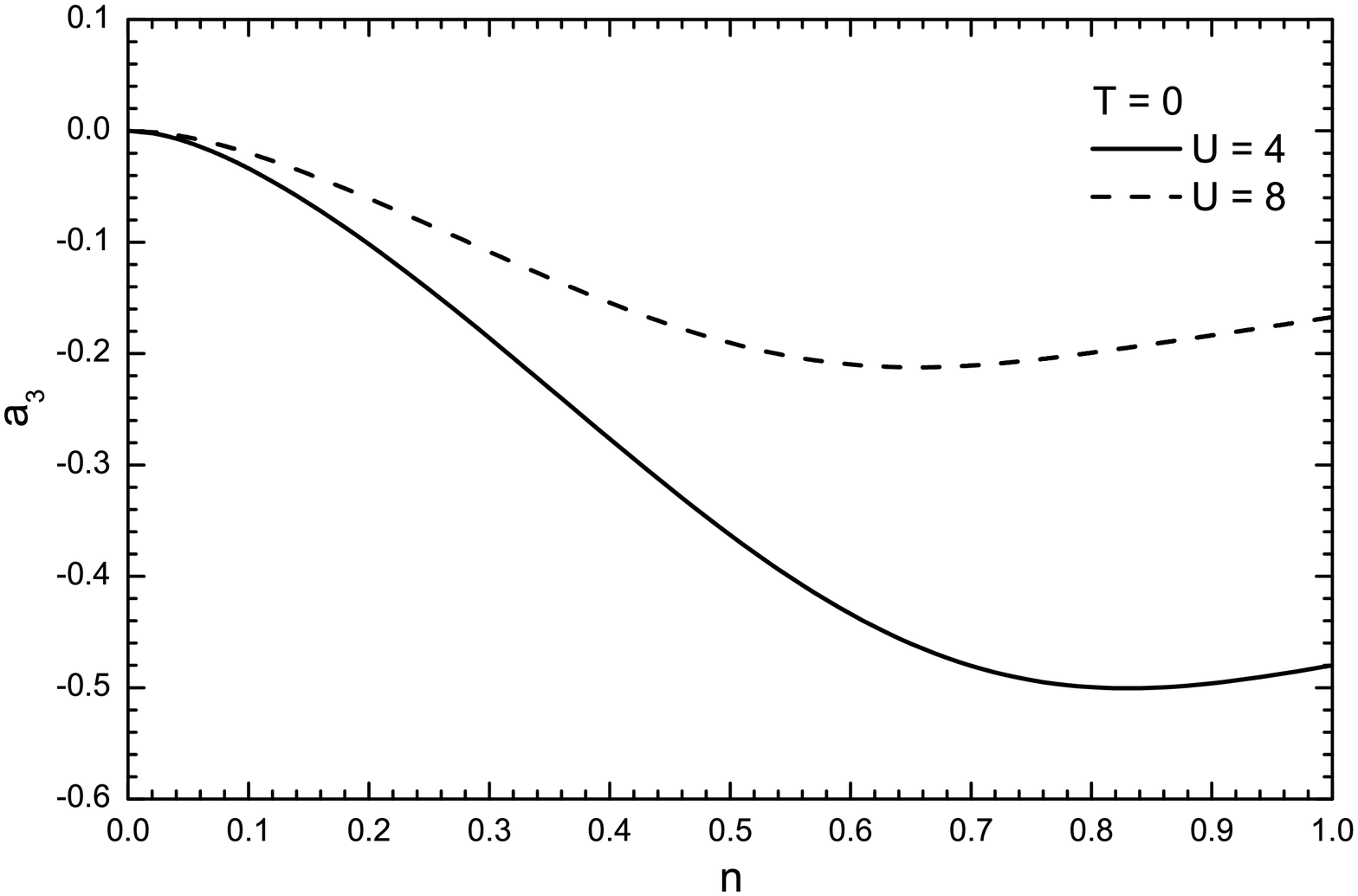}
\includegraphics[width=8cm,keepaspectratio=true]{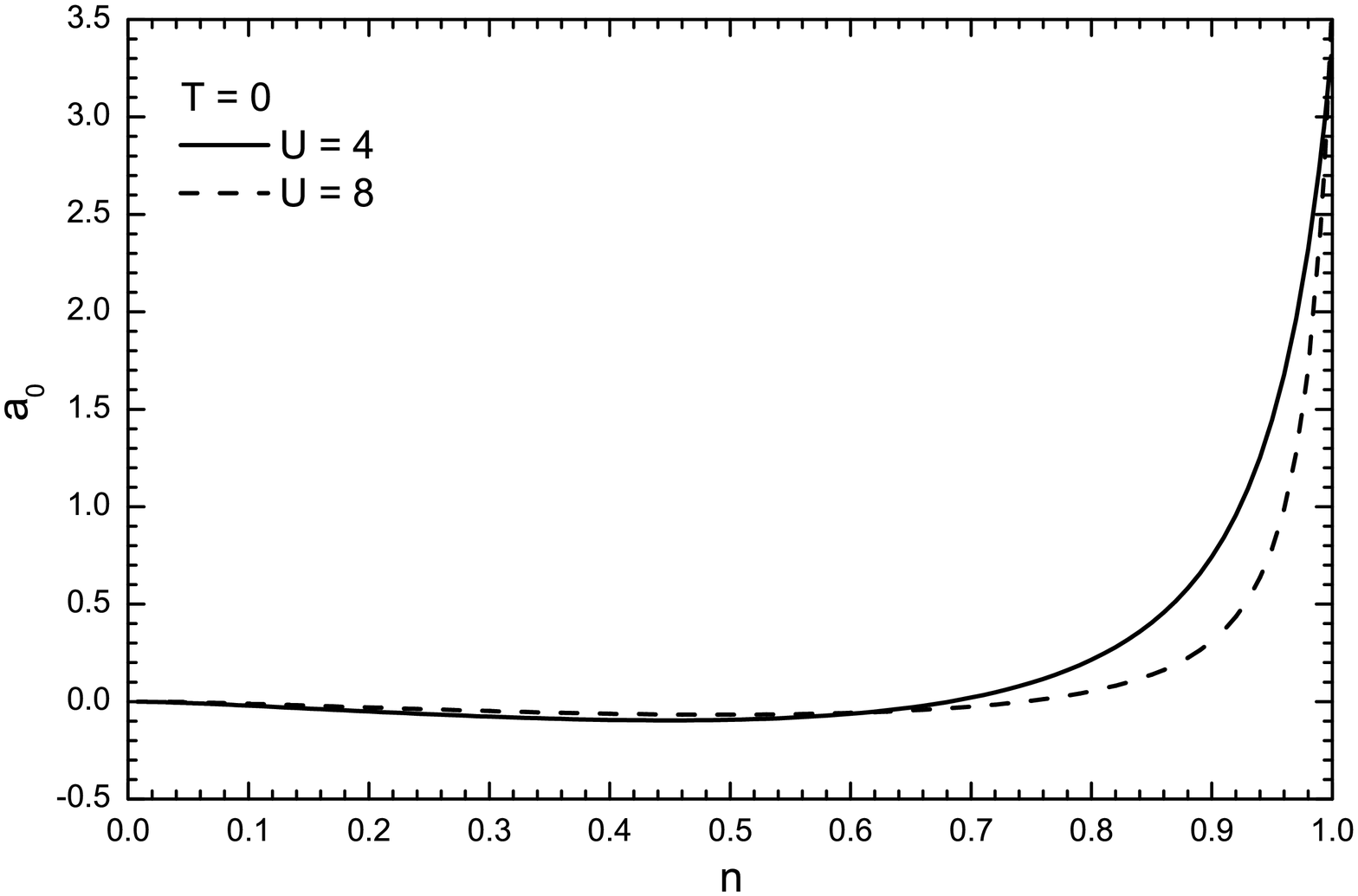}
\end{center}
\caption{$a_3$ and $a_0$ as functions of the filling $n$ for $T=0$
and $U=4$ and $8$.} \label{Fig1}
\end{figure}

Moreover, as the susceptibility has to be single-value at $k=0$ we
have also to require that $m_{3}^{(\mu )}=0$. The coefficients of
$m_{22}^{(\mu )}\left( \mathbf{k}\right)$ in the limit of small
$k$ have the following expressions (see Appendix)
\begin{align}
&m_{0}^{(\mu )} =U(-a_{\mu }+\frac14b_{\mu }+\frac12c_{\mu }+\frac14d_{\mu }) \\
&m_{1}^{(\mu )} =\frac{U}{4}(2a_{\mu }-c_{\mu }-d_{\mu }-2D-2E^{\eta }) \\
&m_{3}^{(\mu )} =-\frac38t(C^{\alpha
}-2C^{\mu }+C^{\lambda }) \nonumber \\
&+\frac{U}{48}(a_{\mu }+c_{\mu }-2d_{\mu }-D+6E^{\beta }-7E^{\eta
})
\end{align}
According to this, we can write the following algebraic relations
for the parameters $ b_{\mu }$, $c_{\mu }$ and $d_{\mu }$
\begin{eqnarray}
b_{\mu } &=&a_{\mu }+3D+E^{\eta
}+2E^{\beta } \nonumber \\
&&-6\frac{t}{U}\left( C^{\alpha }+C^{\lambda }-2C^{\mu }\right) \\
c_{\mu } &=&a_{\mu }-D+E^{\eta
}-2E^{\beta } \nonumber \\
&&+6\frac{t}{U}\left( C^{\alpha }+C^{\lambda }-2C^{\mu }\right) \\
d_{\mu } &=&a_{\mu }-D-3E^{\eta}+2E^{\beta } \nonumber \\
&&-6\frac{t}{U}\left( C^{\alpha }+C^{\lambda }-2C^{\mu }\right)
\end{eqnarray}
and use the Eq. (\ref{Pauli}) to compute the parameter $a_{\mu }$
self-consistently. In Fig.~\ref{Fig1}, we report the behavior of
$a_3$ and $a_0$ as functions of the filling $n$ for $T=0$ and
$U=4$ and $8$. The behavior of $a_3$ reveals a strong dependence
on both filling and Coulomb repulsion of the intensity of spin
correlations. In particular, at half-filling we have the maximum
dependence on $U$. $a_0$, instead, is practically featureless
except for a region near half filling, whose extension is
controlled by the strength of the Coulomb repulsion, where rapidly
and enormously increases with a slope that again depends on $U$.
This latter behavior results in a strong enhancement of the charge
correlations in the proximity of the Mott-Hubbard metal-insulator
transition.

It is necessary to report that this analysis can be considered an
extension and a completion of that done in
Ref.~\onlinecite{Mancini:94a}. The main differences are related to
the use of causal propagator in place of the retarded one and to
the exploitation of the hydrodynamics constraints to fix the
parameters coming in the energy spectra whenever we wish to retain
the complete dependence on the momentum.

\section{Results}

\begin{figure}[tbp]
\begin{center}
\includegraphics[width=8cm,keepaspectratio=true]{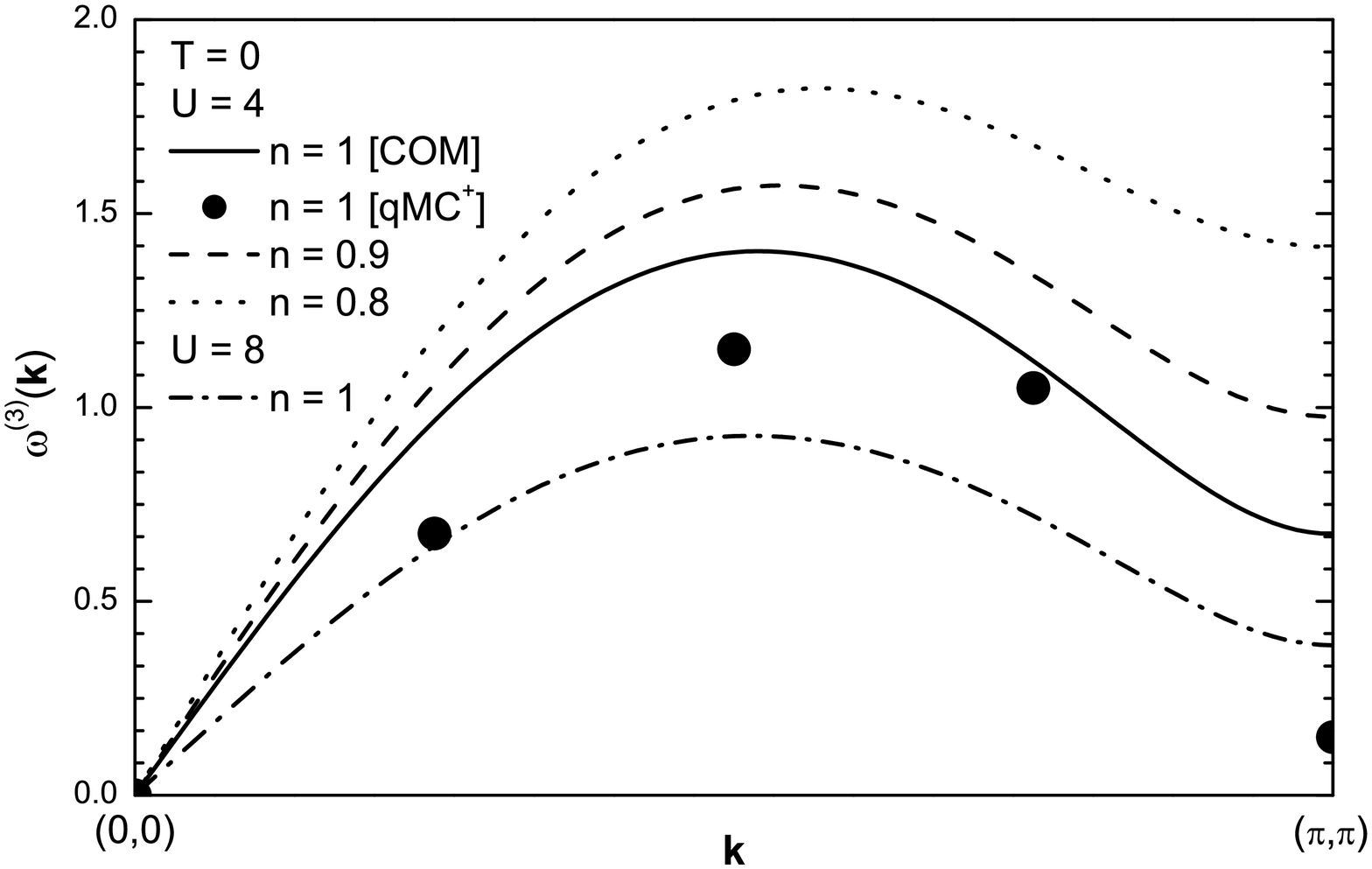}
\includegraphics[width=8cm,keepaspectratio=true]{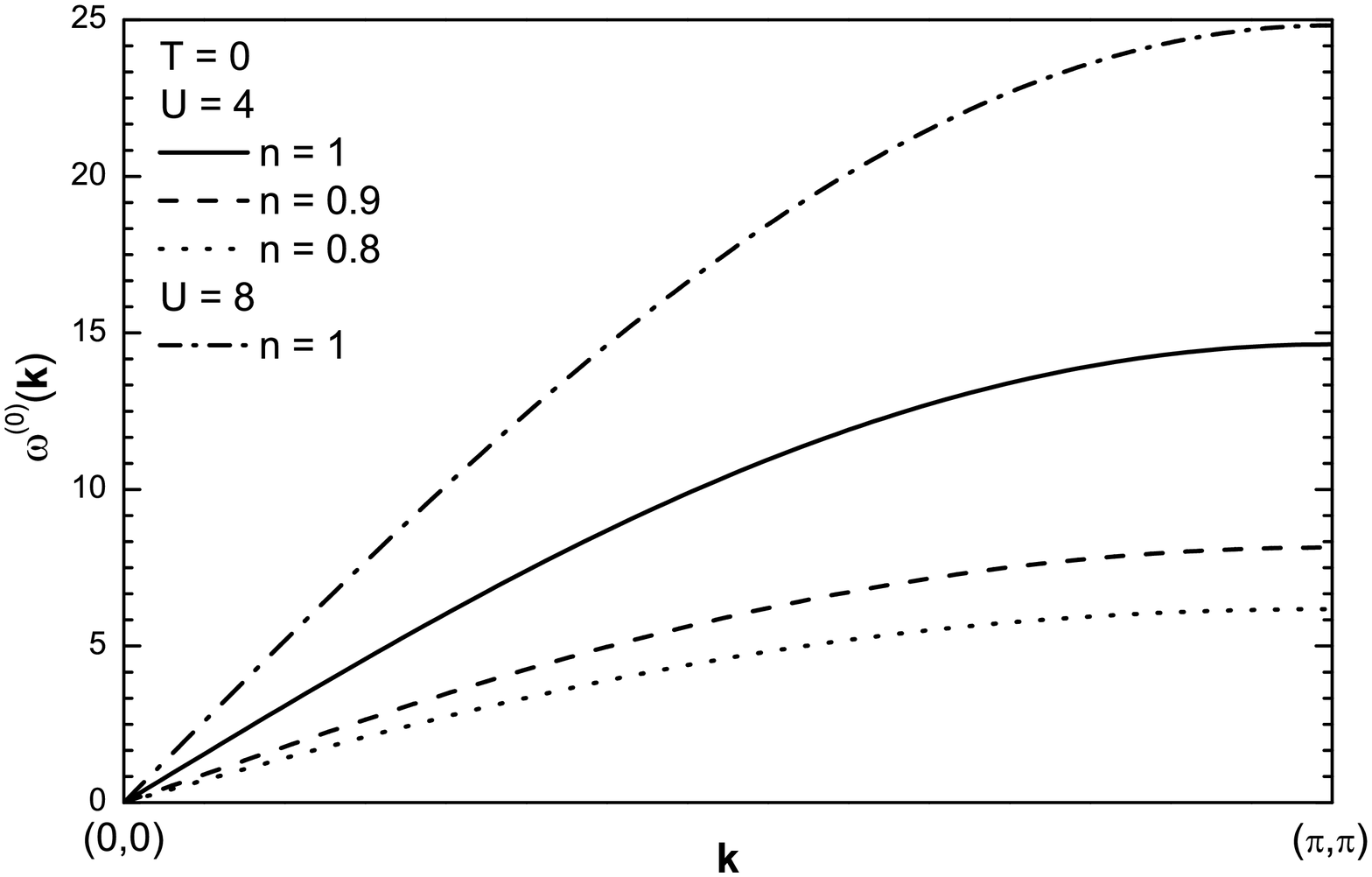}
\end{center}
\caption{The spin $\omega^{(3)}(\mathbf{k})$ and charge
$\omega^{(0)}(\mathbf{k})$ spectra as functions of the momentum
($k_x=k_y$) for $n=1$, $0.9$, $0.8$, $U=4$ and $8$ and $T=0$; the
qMC$^+$ data ($10 \times 10$) for $\omega^{(3)}(\mathbf{k})$ at
$U=4$, $n=1$and $T=0$ are from \citet{White:89}.} \label{Fig2}
\end{figure}

\subsection{Spin and charge spectra}

The spin and charge spectra, as functions of the momentum, are
reported in Fig.~\ref{Fig2} for $n=1$, $0.9$, $0.8$, $U=4$ and $8$
and $T=0$. As regards the spin spectrum, \emph{COM} result is in
fair agreement with the quantum Monte Carlo data\citep{White:89}
($10\times 10$) except for $\mathbf{k}=(\pi ,\pi )=\mathbf{Q}$.
The very small value reported by the numerical data at
$\mathbf{Q}$ can be understood as a consequence of overestimated
antiferromagnetic correlations (i.e., the antiferromagnetic
correlation length actually exceeds the cluster size, see
Fig.~\ref{Fig9}). \emph{COM} results, instead, are obtained with
paramagnetic boundary conditions. The minimum at $\mathbf{Q}$ in
the spin spectrum is the clearest possible evidence that we have
quite strong antiferromagnetic correlations in our solution. The
doping is quite efficient in reducing the intensity of them. On
the contrary, they significatively increase with the Coulomb
repulsion according to the stronger and stronger influence of the
exchange energy $J$ in the strongly interacting regime. The charge
spectrum shows an enhancement of the \emph{velocity} with
decreasing doping and increasing Coulomb repulsion, that is, in
the proximity of a Mott-Hubbard metal-insulator transition, which
would have as signature the divergency of the former.

\begin{figure}[tbp]
\begin{center}
\includegraphics[width=8cm,keepaspectratio=true]{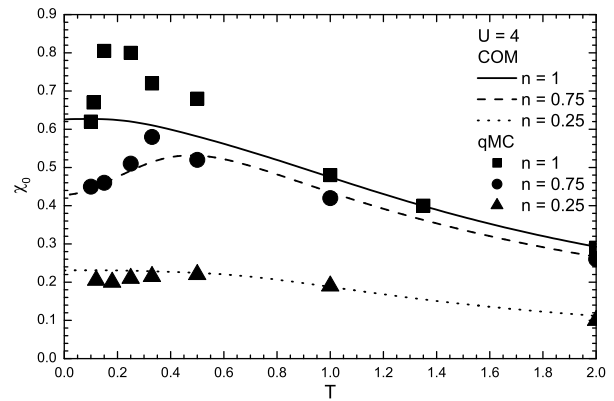}
\end{center}
\caption{The uniform static spin susceptibility $\chi _{0}$ as
function of the temperature $T$ for $U=4$, $n=1$, $0.75$ and
$0.25$; the qMC data ($8 \times 8$) are from \citet{Moreo:93}.}
\label{Fig3}
\end{figure}

\begin{figure}[tbp]
\begin{center}
\includegraphics[width=8cm,keepaspectratio=true]{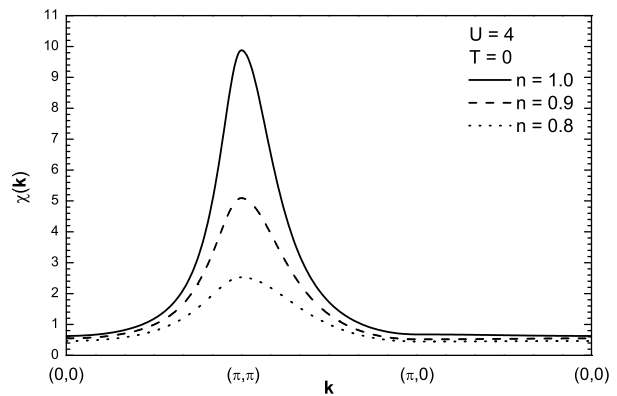}
\end{center}
\caption{The spin susceptibility $\chi(\mathbf{k})$ as function of
the momentum for $U=4$, $T=0$ and $n=1$, $0.9$ and $0.8$.}
\label{Fig4}
\end{figure}

\subsection{Spin susceptibility}

\begin{figure}[tbp]
\begin{center}
\includegraphics[width=8cm,keepaspectratio=true]{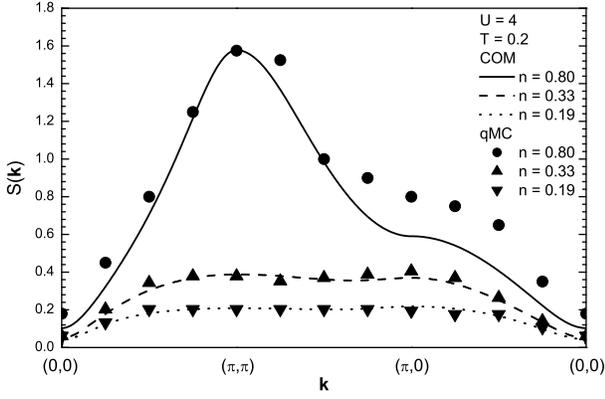}
\end{center}
\caption{The spin correlation function $S(\mathbf{k})$ as function
of the momentum for $U=4$, $T=0.2$ and $n=0.8$, $0.33$ and $0.19$;
the qMC data ($8\times 8$) are from \citet{Vilk:94}.} \label{Fig5}
\end{figure}

\begin{figure}[tbp]
\begin{center}
\includegraphics[width=8cm,keepaspectratio=true]{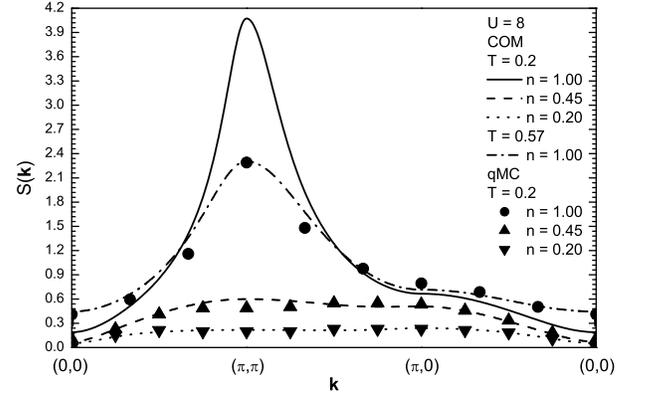}
\end{center}
\caption{The spin correlation function $S(\mathbf{k})$ as function
of the momentum for $U=8$, $T=0.2$ and $T=0.57$ and $n=1$, $0.45$
and $0.2$; the qMC data ($8\times 8$) are from \citet{Vilk:94}.}
\label{Fig6}
\end{figure}

The dynamical spin susceptibility $\chi_s \left( \mathbf{k},\omega
\right)$ can be obtained by Eqs.~(\ref{Rechi}) and (\ref{Imchi})
with $\mu=3$ and has the expression
\begin{equation}
\chi_s \left( \mathbf{k},\omega
\right)=-\sum_{n=1}^{2}\frac{\sigma_{11}^{(n,3)}\left(
\mathbf{k}\right)}{\omega -\omega _{n}^{(3)}\left(
\mathbf{k}\right) +\mathrm{i}\,\delta }
\end{equation}
According to this, the static $\chi\left( \mathbf{k}
\right)=\lim_{\omega \rightarrow 0}\chi_s \left( \mathbf{k},\omega
\right)$ and the static and uniform
$\chi_0=\lim_{\mathbf{k}\rightarrow \mathbf{0}}\chi\left(
\mathbf{k} \right)$ spin susceptibility are given by
\begin{align}
& \chi\left( \mathbf{k} \right) =
\frac{\left\{4\left[1-\alpha\left( \mathbf{k}
\right)\right]C^{\alpha }\right\}^2}{m_{22}^{(3)}\mathbf{k}
} \\
& \chi_0 = -\frac{(4C^{\alpha })^{2}}{24t(C^{\alpha }-C^{\mu
})-U(c_{3}+4E^{\beta }) }
\end{align}
$\chi _{0}$, as a function of the temperature, is reported in
Fig.~\ref{Fig2} for $U=4$ and $n=0.25$, $0.75$ and $1$. \emph{COM}
results are in very good agreement with the quantum Monte Carlo
ones\citep{Moreo:93} ($8\times 8$) for $n=0.25$ and $0.75$. For
$n=1$ and low temperatures, our paramagnetic solution cannot
reproduce the overestimated antiferromagnetic correlations present
in the numerical results. Anyway, our spin susceptibility $\chi
\left( \mathbf{k} \right)$ and our spin correlation function $S
\left( \mathbf{k} \right)$ (see next section) present a large
enhancement at $\mathbf{Q}$ on reducing the doping (see
Fig.~\ref{Fig4}) and increasing the Coulomb repulsion (see
Figs.~\ref{Fig5} and \ref{Fig6}) showing that also \emph{COM}
results present well developed antiferromagnetic correlations
although they should be compatible with the chosen paramagnetic
solution. It is worth noting that the presented results are in
better agreement with quantum Monte Carlo data than the random
phase approximation and the \emph{COM} within the one-loop
approximation (see Ref.~\onlinecite{Mancini:95b} and references
therein).

\subsection{Spin correlation function}

The spin correlation function is defined as
\begin{equation}
S\left( \mathbf{i},\mathbf{j}\right) =\left\langle n_{3}\left(
\mathbf{i},t\right) \,n_{3}\left( \mathbf{j},t\right)
\right\rangle=F^{-1}\left[S\left( \mathbf{k}\right)\right]
\end{equation}
where $F^{-1}$ stands for the Fourier anti-transform and the
structure factor reads as
\begin{equation}
S\left( \mathbf{k}\right)= -\frac{2t\,I^{(3)}_{12}\left(
\mathbf{k} \right)}{\omega^{(3)}\left( \mathbf{k} \right) }\coth
\frac{\omega^{(3)}\left( \mathbf{k}\right) }{2T}
\end{equation}

\begin{figure}[tbp]
\begin{center}
\includegraphics[width=8cm,keepaspectratio=true]{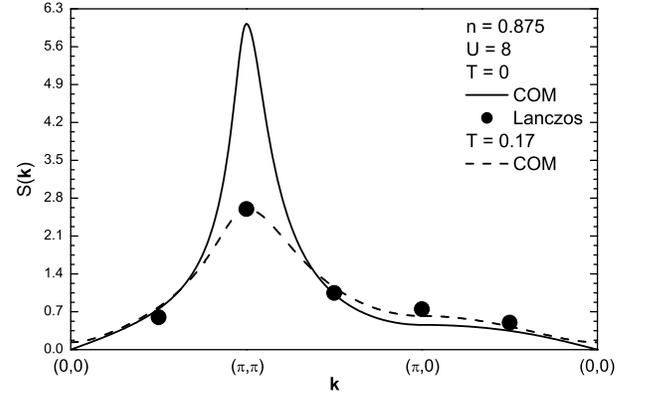}
\end{center}
\caption{The spin correlation function $S(\mathbf{k})$ as function
of the momentum for $n=0.875$, $U=8$ and $T=0$ and $0.17$; the
Lanczos data ($4 \times 4$) at $T=0$ are from \citet{Fano:92}.}
\label{Fig7}
\end{figure}

\begin{figure}[tbp]
\begin{center}
\includegraphics[width=8cm,keepaspectratio=true]{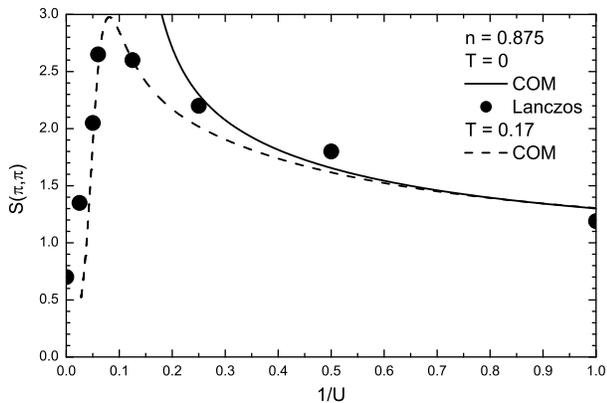}
\end{center}
\caption{The spin correlation function at $\mathbf{Q}$
($S(\mathbf{Q})$) as function of the inverse of $U$ for $n=0.875$
and $T=0$ and $0.17$; the Lanczos data ($4 \times 4$) at $T=0$ are
from \citet{Fano:92}.} \label{Fig8}
\end{figure}

The behavior of $S\left( \mathbf{k}\right)$, as function of the
momentum, is reported in Figs.~\ref{Fig5}, \ref{Fig6} and
\ref{Fig7} in comparison with some numerical
data\citep{White:89,Fano:92,Vilk:94} for different values of
filling, Coulomb repulsion and temperature. We have a very good
agreement with the numerical results for sufficiently high values
of the doping for all shown values of the Coulomb repulsion. In
the proximity of half-filling the numerical data suffer from a
saturation of the antiferromagnetic correlation
length\citep{White:89} that becomes comparable with the size of
the cluster. For $U=4$ and $n=0.8$ (see Fig.~\ref{Fig5}), the
correlation length is slightly smaller than the size of the
cluster: our solution results capable to describe this situation
fairly well (the height of the peak at $\mathbf{Q}$ is exactly
reproduced and the momentum dependence is qualitatively correct,
again practically exact along the diagonal) except for the exact
value of the numerical data along the main axes. This is probably
due again to an overestimation of the correlations by the
numerical analysis owing to the finite size of the cluster. For
$U=8$ and $n=1$ (see Fig.~\ref{Fig6}) and $0.875$ (see
Fig.~\ref{Fig7}), in order to reproduce the numerical data we need
to increase the temperature as to decrease our value of the
correlation length and match that of the numerical analysis, which
is stuck at the saturation value due to the finiteness of the
clusters. The results of such a procedure are astonishing, we
manage to exactly reproduce the numerical data for any value of
the momentum, and not only at $\mathbf{Q}$, revealing the
correctness and power of our approach and the limitations of the
numerical analysis. According to this, we wish to point out that
the numerical data need to be carefully and cleverly interpreted
in order to obtain from them sensible and effective information.
The behavior of the spin correlation function as a function of
$1/U\propto J$\ (the exchange energy) is shown in Fig.~\ref{Fig8}.
Again, in order to obtain results comparable with the numerical
ones\citep{Fano:92} we need to use an higher temperature: at $T=0$
and for high enough values of $U$, our results show a divergency
in the correlation length that is extraneous to the physics of a
finite system. By using the same temperature of Fig.~\ref{Fig7}
(the Lanczos data have the same source), we get a very good
agreement for any regime of interaction: our solution properly
describes also the low-energy dynamics of the spin system.

\begin{figure}[tbp]
\begin{center}
\includegraphics[width=8cm,keepaspectratio=true]{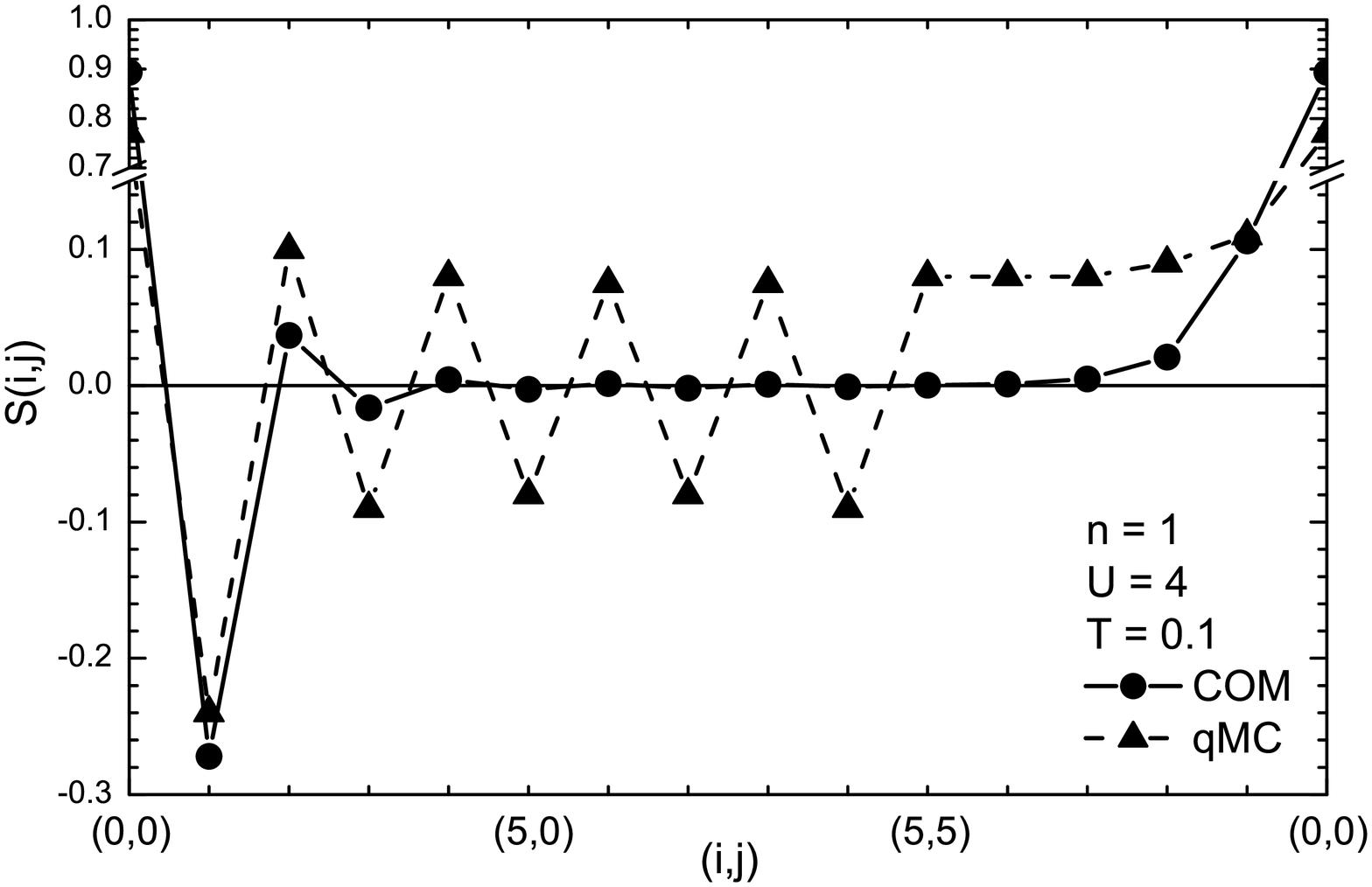}
\includegraphics[width=8cm,keepaspectratio=true]{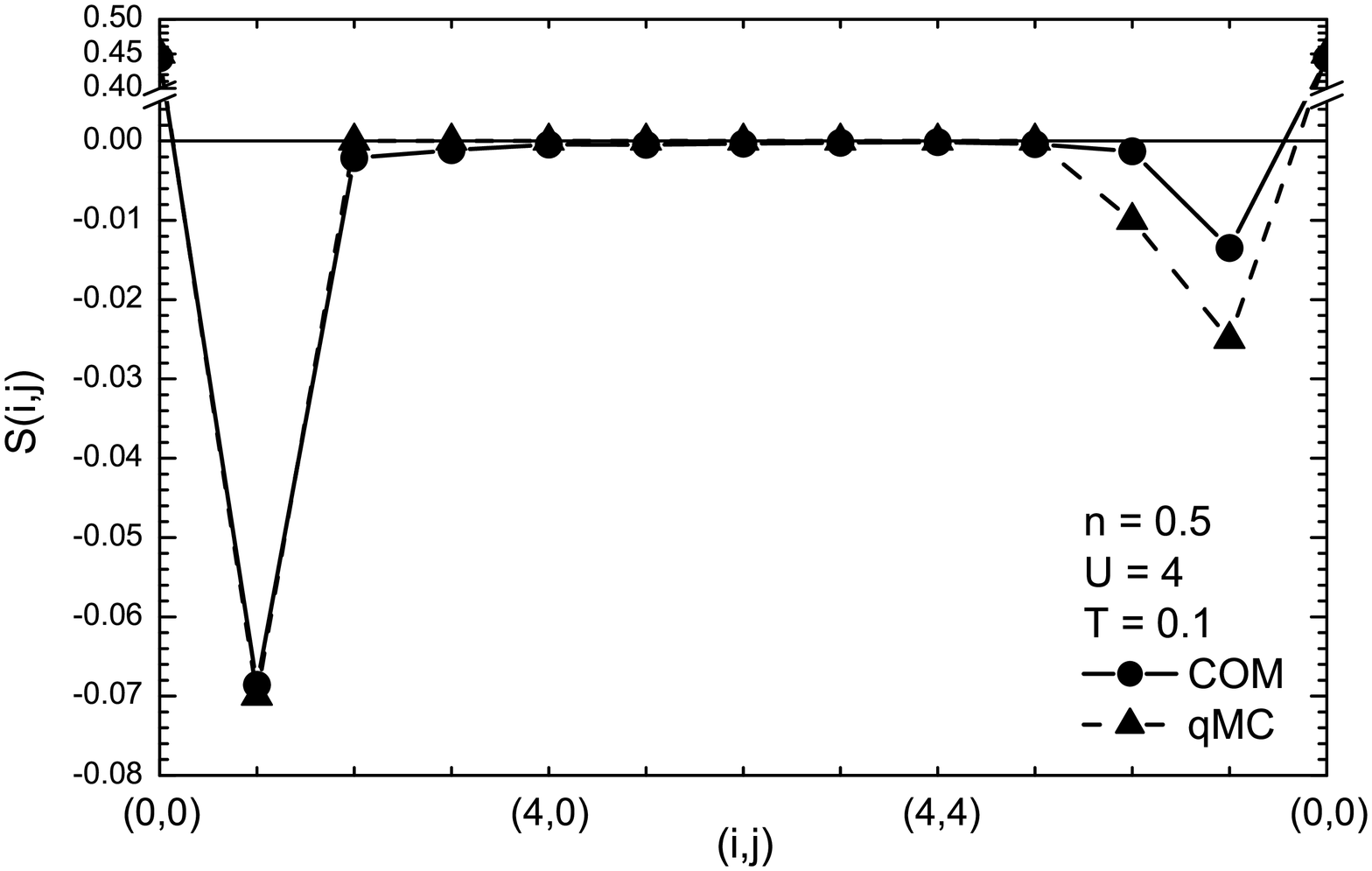}
\end{center}
\caption{The spin correlation function $S(i,j)$ along the
principal directions for $U=4$, $T=0.1$ and (top) $n=1$ [(bottom)
$n=0.5$]; the qMC data ($10 \times 10$) are from
\citet{White:89}.} \label{Fig9}
\end{figure}

In Fig.~\ref{Fig9}, we report the behavior of $S\left( i,j\right)$
along the three principal directions in the lattice for $U=4$,
$T=0.1$ and (top) $n=1$ [(bottom) $n=0.5$]. The quantum Monte
Carlo results\citep{White:89} at $n=1$ present an
antiferromagnetic correlation length as large as the size of the
cluster. The correlation along the principal axes [$
(0,0)\rightarrow (i_{x},0)$ and $(5,0)\rightarrow (5,i_{y})$] is
antiferromagnetic and is ferromagnetic along the diagonals [$
(0,0)\rightarrow (i,i)$] as in an ordinary two-dimensional
Ne\'{e}l phase. \emph{COM} results show exactly the same behavior
although the correlation length is much smaller: we analyze the
paramagnetic phase and for $U=4$ we still not have so well
developed antiferromagnetic correlations. The on-set of an
antiferromagnetic phase (i.e., to have an antiferromagnetic
correlation length as large as the size of the system) for a
finite system seems possible for any finite value of $U$ at
half-filling, while, for an infinite system, it could be related
to the existence of a critical value of the interaction $U$ that,
in our case, falls between $4$ and $8$. Actually, our study of the
antiferromagnetic phase\citep{Avella:00b} confirm that our
critical value is $U_c\cong6$. At $n=0.5$ the agreement becomes
quantitative as the strong antiferromagnetic correlations present
at half filling completely disappear.

\begin{figure}[tbp]
\begin{center}
\includegraphics[width=8cm,keepaspectratio=true]{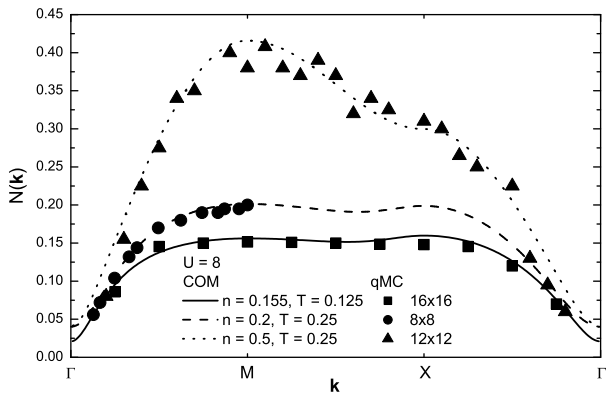}
\end{center}
\caption{The charge correlation function $N(\mathbf{k})$ as a
function of the momentum for $U=8$, $T=0.125$ and $0.25$ and
$n=0.155$, $0.2$ and $0.5$; the qMC data ($8 \times 8$, $12 \times
12$, $16 \times 16$) are from \citet{Chen:94}.} \label{Fig10}
\end{figure}

\begin{figure}[tbp]
\begin{center}
\includegraphics[width=8cm,keepaspectratio=true]{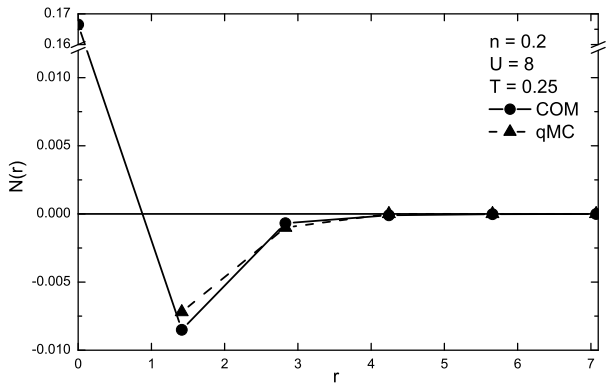}
\end{center}
\caption{The charge correlation function $N(r)$ as a function of
the distance for $n=0.2$, $U=8$ and $T=0.25$; the qMC data ($16
\times 16$) are from \citet{Chen:94}.} \label{Fig11}
\end{figure}

\begin{figure}[tbp]
\begin{center}
\includegraphics[width=8cm,keepaspectratio=true]{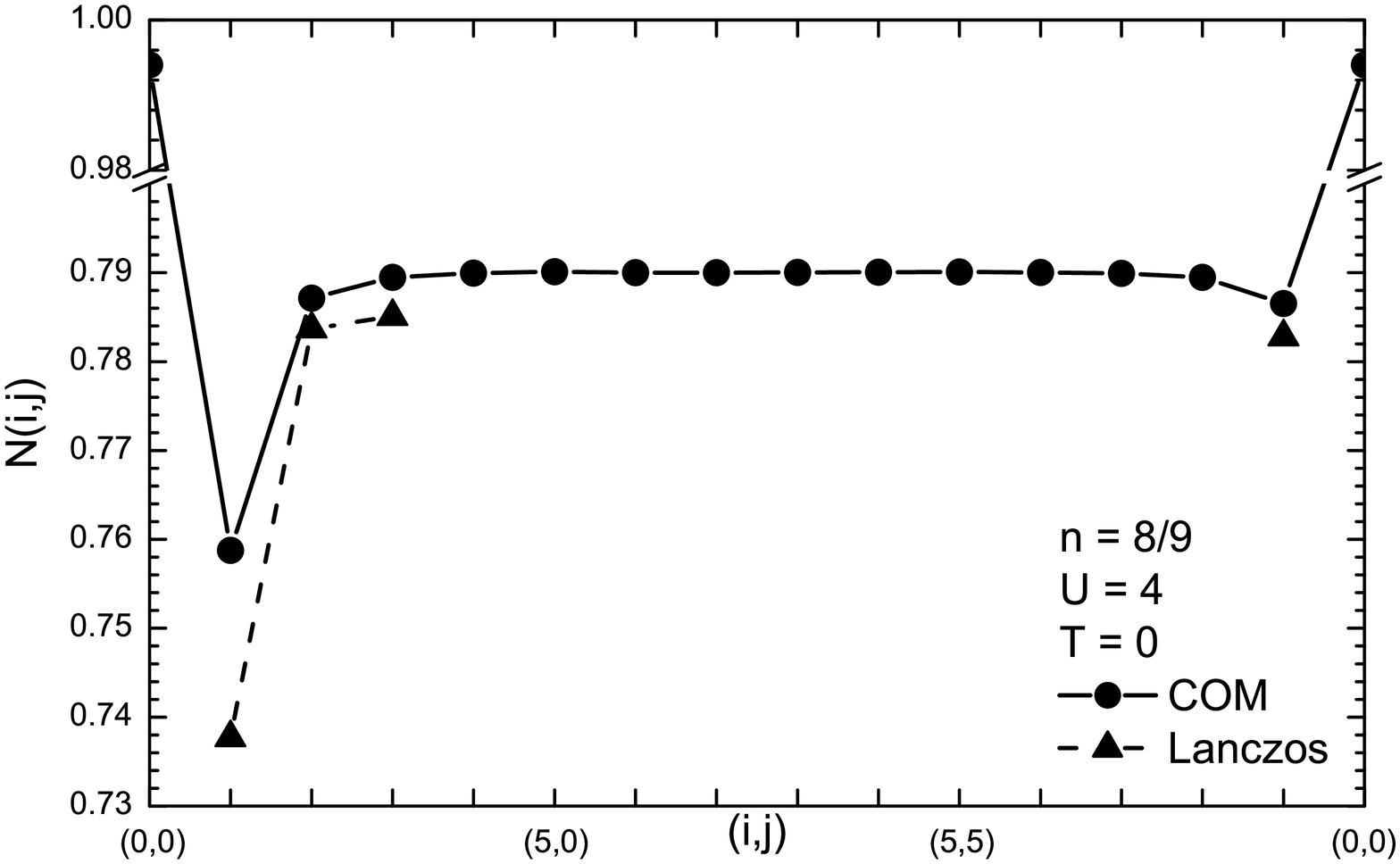}
\includegraphics[width=8cm,keepaspectratio=true]{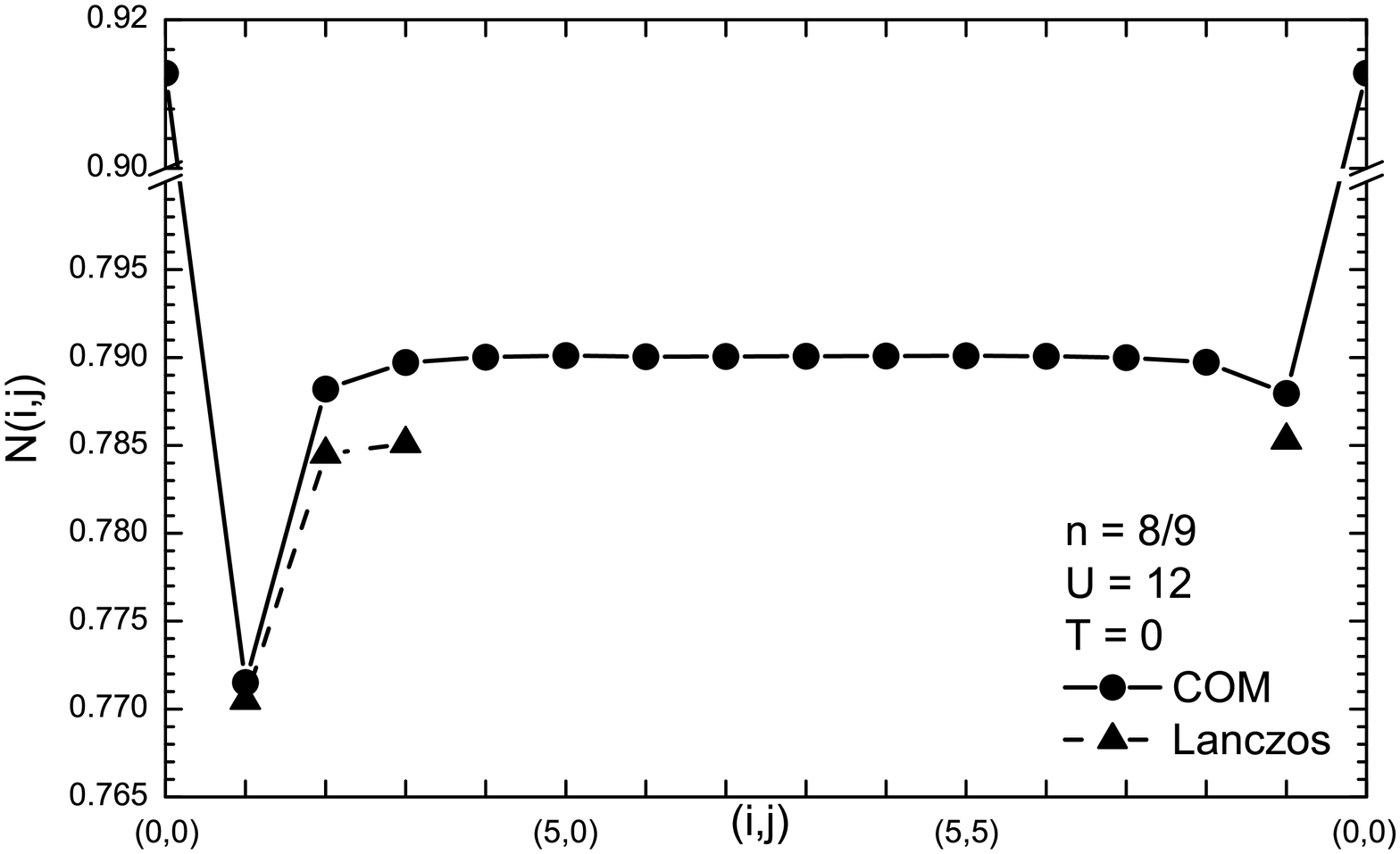}
\end{center}
\caption{The charge correlation function $N(r)$ as a function of
the distance for $n=8/9$, $U=4$ (top) [$U=12$ (bottom)] and $T=0$;
the Lanczos data ($4 \times 4$) are from \citet{Becca:00}.}
\label{Fig12}
\end{figure}

\begin{figure}[tbp]
\begin{center}
\includegraphics[width=8cm,keepaspectratio=true]{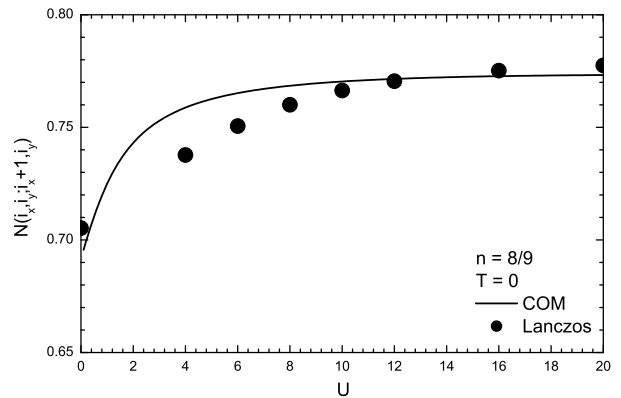}
\end{center}
\caption{The charge correlation function $N\left( i,i^{\alpha
_{x}}\right) $ as a function of $U$ for $n=8/9$ and $T=0$; the
Lanczos data ($4 \times 4$) are from \citet{Becca:00}.}
\label{Fig13}
\end{figure}

\subsection{Charge correlation function}

The charge correlation function is defined as
\begin{equation}
N\left( \mathbf{i},\mathbf{j}\right) =\left\langle n\left(
\mathbf{i},t\right) \,n\left( \mathbf{j},t\right)
\right\rangle=F^{-1}\left[N\left( \mathbf{k}\right)\right]
\end{equation}
where $N\left( \mathbf{k}\right)$ reads as
\begin{equation}
N\left( \mathbf{k}\right)= -\frac{2t\,I^{(0)}_{12}\left(
\mathbf{k} \right)}{\omega^{(0)}\left( \mathbf{k} \right) }\coth
\frac{\omega^{(0)}\left( \mathbf{k}\right) }{2T}
\end{equation}
$N\left( \mathbf{k}\right)$ is reported in Fig.~\ref{Fig10}, as a
function of the momentum, for various fillings and temperatures
and $U=8$. We have again a very good agreement with quantum Monte
Carlo results\citep{Chen:94} for all shown values of the external
parameters and of the momentum. The enhancement at
$\mathbf{k}=\mathbf{Q}=M$ for $ n=0.5$ can be interpreted as the
manifestation of a rather weak ordering of the charge with a
checkerboard pattern. \emph{COM} results manage to reproduce such
double peak structure showing a capability to quantitatively
describe, also in a translational invariant phase, rather strong
charge correlations.

In Figs.~\ref{Fig11} and \ref{Fig12}, we report the behavior of
$N\left( r\right) $, as a function of the distance
$r=\sqrt{i^{2}+j^{2}}$, for $U=4$ and $12$, $T=0.01$ and $ n=8/9$
and $U=8$, $T=0.25$ and $n=0.2$, respectively. \emph{COM} results
are in good quantitative agreement with the numerical
results\citep{Chen:94,Becca:00} showing once more that the charge
dynamics is really well described by our solution. In
Fig.~\ref{Fig13}, $N\left( i,i^{\alpha _{x}}\right) $ is shown as
a function of the Coulomb repulsion $U$ for $n=8/9$ and $T=0$. The
agreement with Lanczos data\citep{Becca:00} is quite good and gets
better and better as $U$ increases.

\section{Conclusions}

An analytical description of the charge and spin dynamics of the
two-dimensional Hubbard model in the paramagnetic phase has been
presented within a two-pole approximation in the framework of the
\emph{COM}. The hydrodynamics constraints as well as the Pauli
principle requirements have been embedded in the fully
self-consistent solution by the very beginning and any decoupling
has been avoided. The antiferromagnetic correlations are really
well described together with some weak charge ordering tendency at
commensurate filling. Spin spectrum, static uniform spin
susceptibility, spin and charge correlation functions are in very
good agreement with the numerical results present in the
literature and clearly state the reliability of the proposed
procedure.

\begin{acknowledgments}
V.T. wishes to thank all the members of the Dipartimento di Fisica
``E.R. Caianiello'', Universit\`{a} degli Studi di Salerno, and
especially Prof.~F.~Mancini, for the kind hospitality.
\end{acknowledgments}

\appendix

\section*{Appendix}

We have the following expressions for the $m^{(\mu)}$-matrix
entries
\begin{eqnarray}
I_{l_{\mu }\rho _{\mu }}\left( \mathbf{k}\right) &=& \frac34\left[
1-\alpha \left( \mathbf{k}\right)
\right]\left(12C^{\alpha}+C^{\lambda}+6C^{\mu}\right) \nonumber \\
&&-\frac34\left[ 1-\eta \left( \mathbf{k}\right) \right]\left(
C^{\alpha }+C^{\lambda }+2C^{\mu
}\right) \nonumber \\
&&+\frac14\left[ 1-\lambda \left( \mathbf{k}\right)
\right]C^{\lambda }
+\frac32\left[ 1-\mu \left( \mathbf{k}\right) \right]C^{\mu } \nonumber \\
&& -3\left[ 1-\beta \left( \mathbf{k}\right) \right]\left(
C^{\alpha }+C^{\mu
}\right) \\
I_{\kappa_{\mu }\rho _{\mu }}\left( \mathbf{k}\right) &=&-2\left[
1-\alpha
\left( \mathbf{k}\right) \right] D \nonumber \\
&& + \left[ 1-2\alpha
\left( \mathbf{k}\right) \right]\left(2E^\beta+E^\eta\right) \nonumber \\
&&+\eta \left( \mathbf{k}\right) \,E^{\eta }+2\beta \left(
\mathbf{k}\right)
\,E^{\beta } \nonumber \\
&&+\left[ 1-2\alpha \left( \mathbf{k}\right) \right] a_{\mu } \nonumber \\
&&+\frac14\left[b_{\mu }+2\beta \left( \mathbf{k}\right) \,c_{\mu
}+\eta \left( \mathbf{k} \right) \,d_{\mu }\right]
\end{eqnarray}
The following definitions have been used
\begin{eqnarray}
C^{\alpha } &=&\left\langle c^{\alpha }\left( i\right)
\,c^{\dagger }\left(
i\right) \right\rangle \\
%C^{\alpha ^{3}} &=&\left\langle c^{\alpha ^{3}}\left( i\right)
%\,c^{\dagger
%}\left( i\right) \right\rangle \\
C^{\lambda } &=&\left\langle c^{\lambda }\left( i\right)
\,c^{\dagger
}\left( i\right) \right\rangle \\
C^{\mu } &=&\left\langle c^{\mu }\left( i\right) \,c^{\dagger
}\left(
i\right) \right\rangle \\
%E &=&\left\langle c\left( i\right) \,\eta ^{\dagger }\left(
%i\right)
%\right\rangle \\
%E^{\alpha ^{2}} &=&\left\langle c^{\alpha ^{2}}\left( i\right)
%\,\eta
%^{\dagger }\left( i\right) \right\rangle \\
E^{\beta } &=&\left\langle c^{\beta }\left( i\right) \,\eta
^{\dagger
}\left( i\right) \right\rangle \\
E^{\eta } &=&\left\langle c^{\eta }\left( i\right) \,\eta
^{\dagger }\left( i\right) \right\rangle
\end{eqnarray}
\begin{eqnarray}
a_{\mu } &=&2\left\langle c^{\dagger }\left( i\right) \,\sigma
_{\mu }\,c^{\alpha }\left( i\right) \,c^{\dagger }\left( i\right)
\,\sigma _{\mu }\,c^{\alpha }\left( i\right) \right\rangle \nonumber \\
&&-\left\langle c^{\alpha \dagger }\left( i\right) \,\sigma _{\mu
}\,\sigma ^{\lambda }\,\sigma _{\mu }\,c^{\alpha }\left( i\right)
\,n_{\lambda }\left(
i\right) \right\rangle \\
b_{\mu } &=&2\left\langle c^{\dagger }\left( i\right) \,\sigma
_{\mu }\,c^{\dagger }\left( i\right) \,\sigma _{\mu }[c\left(
i\right)
\,c\left( i\right) ]^{\alpha }\right\rangle \nonumber \\
&&-\left\langle c^{\dagger }\left( i\right) \,\sigma _{\mu
}\,\sigma ^{\lambda }\,\sigma _{\mu }\,c\left( i\right)
\,n_{\lambda }^{\alpha }\left(
i\right) \right\rangle \\
c_{\mu } &=&2\left\langle c^{\dagger }\left( i\right) \,\sigma
_{\mu }\,c^{\dagger }(i^{\eta })\,\sigma _{\mu }\,c(i^{\alpha
})\,c(i^{\alpha })\right\rangle \nonumber \\
&&-\left\langle c^{\dagger }\left( i\right) \,\sigma _{\mu
}\,\sigma ^{\lambda }\,\sigma _{\mu }\,c(i^{\eta })\,n_{\lambda
}(i^{\alpha
})\right\rangle \\
d_{\mu } &=&2\left\langle c^{\dagger }\left( i\right) \,\sigma
_{\mu }\,c^{\dagger }(i^{\beta })\,\sigma _{\mu }\,c(i^{\alpha
})\,c(i^{\alpha })\right\rangle \nonumber \\
&&-\left\langle c^{\dagger }\left( i\right) \,\sigma _{\mu
}\,\sigma ^{\lambda }\,\sigma _{\mu }\,c(i^{\beta })\,n_{\lambda
}(i^{\alpha })\right\rangle
\end{eqnarray}
where we used the notation
\begin{eqnarray}
i &=&(i_{x},i_{y},t) \\
i^{\alpha } &=&(i_{x}+a,i_{y},t) \\
i^{\eta } &=&(i_{x}+2a,i_{y},t) \\
i^{\beta } &=&(i_{x}+a,i_{y}+a,t)
\end{eqnarray}

The functions $\beta _{ij}$, $\eta _{ij}$, $\mu _{ij}$ and
$\lambda _{ij}$, the projectors on the second, third, fourth and
fifth nearest neighbors, respectively, have the following
expressions in momentum space
\begin{eqnarray}
\beta \left( \mathbf{k}\right) &=&\frac{1}{2}\left\{ \cos \left[
a\left( k_{x}+k_{y}\right) \right] +\cos \left[ a\left(
k_{x}-k_{y}\right) \right]
\right\} \\
\eta \left( \mathbf{k}\right) &=&\frac{1}{2}\left[ \cos \left(
2a\,k_{x}\right) +\cos \left( 2a\,k_{y}\right) \right] \\
\mu \left( \mathbf{k}\right) &=&\frac{1}{4}\left\{ \cos \left[
a\left( 2k_{x}+k_{y}\right) \right] +\cos \left[ a\left(
k_{x}+2k_{y}\right) \right]
\right. \nonumber \\
&&\left. +\cos \left[ a\left( 2k_{x}-k_{y}\right) \right] +\cos
\left[ a\left( k_{x}-2k_{y}\right) \right] \right\} \\
\lambda \left( \mathbf{k}\right) &=&\frac{1}{2}\left[ \cos \left(
3a\,k_{x}\right) +\cos \left( 3a\,k_{y}\right) \right]
\end{eqnarray}
The following relations hold
\begin{eqnarray}
c^{\alpha ^{2}}\left( i\right) &=&\frac{1}{4}\left[ c\left(
i\right)
+2c^{\beta }\left( i\right) +c^{\eta }\left( i\right) \right] \\
c^{\alpha ^{3}}\left( i\right) &=&\frac{1}{16}\left[ 9c^{\alpha
}\left( i\right) +c^{\lambda }\left( i\right) +6c^{\mu }\left(
i\right) \right]
\end{eqnarray}

\bibliographystyle{apsrev}
\bibliography{Biblio}

\end{document}